\documentclass[english]{iopart}
\usepackage[T1]{fontenc}
\usepackage[latin1]{inputenc}
\usepackage{verbatim}
\usepackage{graphicx}
\usepackage{amssymb}

\makeatletter

\usepackage{babel}
\makeatother

\begin{document}

\newcommand{\avg}[1]{\langle#1\rangle}
\newcommand{\p}{\prime}
\newcommand{\dg}{\dagger}
\newcommand{\ket}[1]{|#1\rangle}
\newcommand{\bra}[1]{\langle#1|}
\newcommand{\proj}[2]{|#1\rangle\langle#2|}
\newcommand{\pd}[2]{\frac{\partial#1}{\partial#2}}
\newcommand{\der}[2]{\frac{d #1}{d #2}}

\title[Numerical ansatz for integro-differential equations]{Numerical ansatz for solving integro-differential equations with
increasingly smooth memory kernels: spin-boson model and beyond}
\author{Michael Zwolak}
\address{Institute for Quantum Information, California Institute of Technology,
Pasadena, California 91125}
\ead{zwolak@theory.caltech.edu}

\begin{abstract}
We present an efficient and stable numerical ansatz for solving a 
class of integro-differential equations. We define the class as integro-differential
equations with \emph{increasingly smooth} memory kernels. The resulting
algorithm reduces the computational cost from the usual $T^{2}$ to
$TC\left(T\right)$, where $T$ is the total simulation time and $C\left(T\right)$
is some function. For instance, $C\left(T\right)$ is equal to $\ln T$
for polynomially decaying memory kernels. Due to the common occurrence
of increasingly smooth memory kernels in physical, chemical, and biological
systems, the algorithm can be applied in quite a wide variety of situations.
We demonstrate the performance of the algorithm by examining two cases.
First, we compare the algorithm to a typical numerical procedure for
a simple integro-differential equation. Second, we solve the NIBA
equations for the spin-boson model in real time.
\end{abstract}
\maketitle

\section{Introduction}

A major problem in the study of quantum systems is the influence of
the environment on its dynamics. A variety of techniques, such as
the influence functional \cite{feynman},  nonequilibrium perturbation theory \cite{keldysh,kadanoff},
and the weak-coupling approximation \cite{breuer}, have been developed
to address this issue. The general approach is to derive equations
of motion which only involve the degrees of freedom of the system.
When working with systems weakly coupled to a rapidly relaxing environment,
one can make the Markov approximation, which results in a local in
time, first-order differential equation. This Markovian master equation
is relatively easy to solve either analytically or numerically. As soon as
the restrictions of weak coupling or fast environmental relaxation
are removed, however, integro-differential equations appear, which
reflects the fact that the environment retains a memory of the system
at past times. 

In particular, a number of interesting physical systems have integro-differential
equations of motion with polynomially decaying memory kernels. For
instance, the spin-boson model \cite{leggett} and Josephson Junctions
with quasi-particle dissipation or resistive shunts \cite{schon}
have $1/t^{2}$ asymptotic kernels when going through a dissipatively
driven phase transition. The occurrence of polynomially decaying kernels
is common in physics and chemistry because many open systems are connected
to an ohmic environment. Polynomially decaying kernels appear in biological
applications as well. \cite{cushing} Due to the frequent appearance
of this type of kernel and also others which satisfy a requirement
we call \emph{increasingly smooth} below, it would be beneficial to
have efficient numerical methods to handle the corresponding integro-differential
equations. Such a method will be useful in applications in many disciplines.
In addition, the method will enable the simulation of memory-dependent
dissipation when used in conjunction with other computational methodologies,
such as matrix product state algorithms for open systems. \cite{zwolak,verstraete,vidal1,vidal2}

In this paper, we give an efficient and stable numerical ansatz for
solving integro-differential equations with increasingly smooth memory
kernels. Using typical techniques, one incurs a computational cost
of $AT^{2}\propto N^{2}$, where $A$ is some constant, $T$ is the
total simulation time, and $N$ is the number of time steps in the
simulation. Thus, there is a considerable advantage in using a high-order
approach to reduce the required value of $N$ as much as possible.
However, the computational cost of the algorithm described herein
scales as $TC\left(T\right)$, where
$C\left(T\right)$ is some function that depends on the form of the
memory kernel. For example, $C\left(T\right)$ is equal to $\ln T$
for polynomially decaying kernels. Such a reduction represents a substantial
advancement in numerically solving integro-differential equations
since it allows for the efficient calculation of long-time behaviour. 

This paper is organized as follows: in section \ref{sec:Algorithm},
we introduce the types of equations under consideration, define {\em increasingly
smooth}, and present the numerical ansatz with a discussion of errors.
We demonstrate the performance of the algorithm using two example
integro-differential equations in section \ref{sec:Examples}. First,
we consider an integro-differential equation composed of a polynomially
decaying kernel and an oscillating function. Comparing directly to
a two stage Runge-Kutta method, we see a large improvement in the scaling
of the error as a function of computational cost. Second, we solve
the noninteracting blip approximation (NIBA) equations for the spin-boson
model in real time. We show that one can rapidly obtain the full real
time solution. In \ref{sec:high}, we discuss how to extend the algorithm to 
higher orders. In \ref{sec:rk}, we outline the Runge-Kutta method
we use as a comparison to the numerical ansatz. In \ref{sec:NIBA}, we
give a simple derivation of the NIBA equations, showing the physical situation 
behind the appearance of an increasingly smooth memory kernel.

\section{Algorithm}
\label{sec:Algorithm}

We want to be able to numerically solve linear integro-differential
equations of the form 
\begin{eqnarray}
\pd{\rho\left(t\right)}{t} & = &
 -\mathcal{K}\int_{0}^{t}dt^{\p}\alpha\left(t-t^{\p}\right)e^{-\mathcal{L}\left(t-t^{\p}\right)}\mathcal{K}^{\p}\rho\left(t^{\p}\right) \\
& = & 
 -\mathcal{K}\int_{0}^{t}d\tau \alpha\left(\tau\right)e^{-\mathcal{L}\left(\tau\right)}\mathcal{K}^{\p}\rho\left(t-\tau\right)
\label{eq:integro}
\end{eqnarray}
where $\rho(t)$ is some quantity of interest (like a density matrix) and $\alpha(\tau)$ is the memory kernel. 
$\mathcal{K}$, $\mathcal{K}^{\p}$, and $\mathcal{L}$ are time-independent operators. An equation of the
form \ref{eq:integro} appears in open quantum mechanical systems in the Born approximation to the full
master equation \cite{breuer}, some exact non-Markovian master equations \cite{zwolak2},
and in phenomenological memory kernel master equations.\cite{lidar,barnett,daffer}
For an arbitrary form of the memory kernel, it is necessary
to compute the integral on the right-hand side of \ref{eq:integro}
at each time step in the numerical simulation. Thus, ignoring the error of the simulation, the computational
cost scales as $T^{2}$, 
where $T$ is the total simulation time. This is
prohibitively expensive in all but the shortest time simulations.

On the opposite extreme is when the memory kernel has an exponential
form 
\begin{equation}
\alpha(t-t^{\p})=\gamma e^{-\gamma\left(t-t^{\p}\right)} \, .
\label{eq:expmem}
\end{equation}
In this case, both functions responsible for evolving the integrand 
($\alpha\left(t-t^{\p}\right),e^{\mathcal{L}\left(t-t^{\p}\right)}$) 
obey simple differential equations in $t$. This allows us to define
a {}``history'' 
\begin{equation}
H(t)=\int_{0}^{t}dt^{\p}\alpha(t-t^{\p})e^{\mathcal{L}(t-t^{\p})}\mathcal{K}^{\p}\rho(t^{\p})
\label{eq:history}
\end{equation}
which obeys the differential equation 
\begin{equation}
\dot{H}(t)=\gamma\mathcal{K}^{\p}\rho(t)-\gamma H(t)-\mathcal{L}H(t) \, .
\label{eq:solexp1}
\end{equation}
By solving this local in time differential equation together with
\begin{equation}
\dot{\rho}(t)=-\mathcal{K}H(t) \, ,
\label{eq:solexp2}
\end{equation}
we can solve the integro-differential equation with a computational
cost scaling as $T$. Cases in between these two extremes have a spectrum
of computational costs scaling from $T$ to $T^{2}$.

\subsection{Increasing Smoothness}

We are interested in integro-differential equations of the form \ref{eq:integro}
with memory kernels which are \emph{increasingly smooth} 
\begin{equation}
\pd{}{t}\left|\frac{\alpha^{\p}\left(t\right)}{\alpha\left(t\right)}\right|<0\,.
\label{eq:incsmooth}
\end{equation}
The idea behind increasing smoothness will become clear below.
More specifically in this paper, we will look at examples where the
memory kernel decays polynomially outside some cut-off time \begin{equation}
\alpha(t)=\left\{ \begin{array}{cc}
\alpha_{\Delta T}\left(t\right) & t<\Delta T\\
\propto\frac{1}{t^{p}} & t\geq\Delta T\end{array}\right.\label{eq:poly}\end{equation}
where $\alpha_{\Delta T}\left(t\right)$ is a bounded but otherwise
arbitrary function and $p$ is some positive real number.\footnote{For the algorithm, 
$p$ need not be greater than one. However, physical equations of motion, 
such as the NIBA equations at $\alpha=1/2$, have an additional factor in the 
kernel to ensure its integral is bounded.} $\Delta T$
is a cut-off which allows us to ignore any transient behaviour in the
memory kernel which does not satisfy the increasingly smooth condition
\ref{eq:incsmooth}. There will generally be some natural cut-off
to the integral. For the overall scaling of the computational cost
for large simulation times, we can ignore the resulting problem-dependent,
constant computational cost due to the integral at times $t<\Delta T$.

Below we will give an algorithm that is specific to the polynomial
kernel \ref{eq:poly}. Similar algorithms are possible for any kernel
which gets smoother as the time argument gets larger. The computational
speedup will depend on how the function gets smoother. For instance,
the polynomially decaying kernel gets smoother as \begin{equation}
\left|\frac{\pd{1/t^{p}}{t}}{1/t^{p}}\right|=\frac{p}{t}\label{eq:smrat}\end{equation}
which allows one to take larger integration blocks as $t$ gets larger.
In this case, one can take a grid with spacing $\delta\propto t$ to cover the integration region. We call 
the partition around a grid point a {\em block}. In this case, the number of blocks scales logarithmically with the total
time of the simulation. Consider another form of the kernel, $\beta\left(t\right)$,
that gets smoother even faster than polynomial, for example 
\begin{equation}
\left|\frac{\beta^{\p}\left(t\right)}{\beta\left(t\right)}\right|=be^{-ct}\ .
\label{eq:faster}
\end{equation}
This will give an even smaller number of blocks (that have exponential size in $t$)
needed to cover a given range of integration and thus 
a faster speedup.\footnote{\ref{eq:faster} has two solutions. Only the one with bounded integral would be physically relevant.} 
In this case, though, a simple truncation of the integral is 
sufficient, which is not the case for a polynomially decaying
kernel.

\subsection{Blocking algorithm for a polynomially decaying kernel}
\label{sub:Blocking-Algorithm}

The algorithm is designed to take advantage of the fact that at each
new time step in solving the integro-differential equation \ref{eq:integro},
we almost already have the integral on the right hand side. Thus,
if we have $\rho\left(t\right)$ at all the previous time steps and
we group them together in a particular way, we should be able to do away
with both storing all these previous values and the need for a full
integration of the right hand side. 
In more concrete terms, we want to be
able to evaluate the history integral 
\begin{equation}
I\left(T,\Delta T\right)=\int_{\Delta T}^{T}d\tau\alpha\left(\tau\right)F\left(\tau,T\right) \, ,
\label{eq:theint}
\end{equation}
with 
\begin{equation}
F\left(\tau,T\right)=e^{\mathcal{L}\tau}\mathcal{K}^{\p}\rho(T-\tau) \, ,
\end{equation}
in such a way that when $T\to T+h$ in the integro-differential equation
\ref{eq:integro}, all the pieces used to compute the integral can
be recycled by evolving for a time step $h$ and then added back together
to get the new integral. 
This is just what is achieved for the exponential
memory kernel \ref{eq:history}, but in that case it is very extreme,
the old integral simply has to be multiplied by a factor and added
to a new piece. Below we show how to evaluate the integral with a
blocking scheme that requires only $\ln T$ number of blocks to cover 
the integration region, can
achieve any desired accuracy, and has blocks that can all be recycled
when $T\to T+h$. Again, we emphasize that the scheme we
give now is specific for polynomially decaying memory kernels. The
basic scheme can be constructed for other memory kernels as well simply
by choosing a block size such that the block size times the smoothness
stays constant.

\begin{figure}
\begin{center}
\includegraphics[width=7cm]{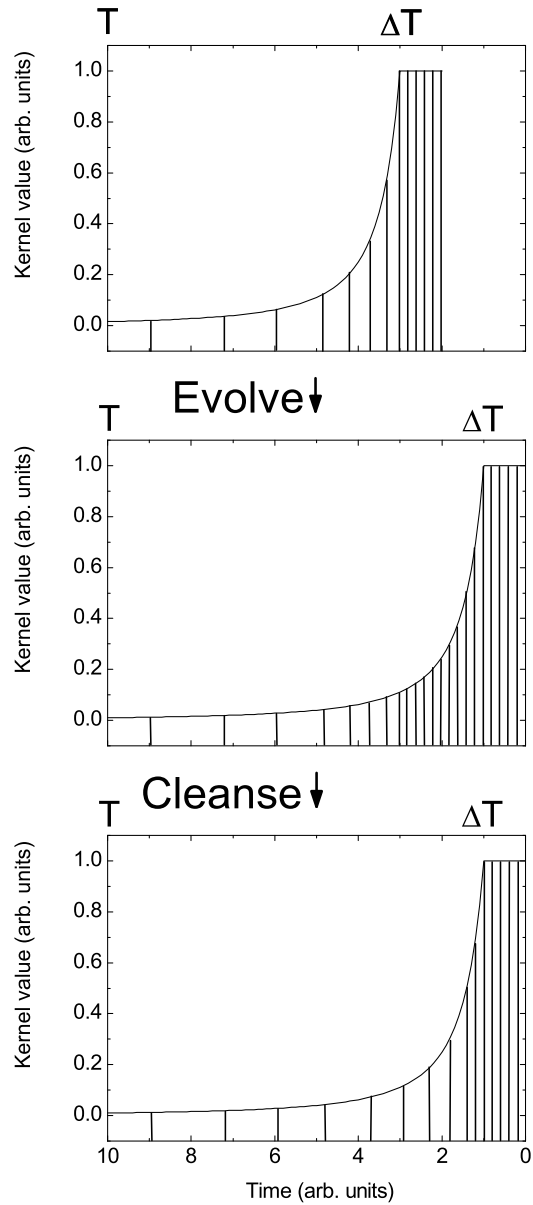}
\end{center}
\caption{Procedure for evolving the integral \ref{eq:theint} in time. We
start with the integral represented as a sum over variable size blocks,
with short times having constant block sizes and longer times having
larger block sizes. We evolve for some amount of time (here shown
as a large amount of time, but in practice this whole procedure is
done discretely in the small time steps of the differential equation).
Then we cleanse the blocks, grouping them together so long as their
size is not too big for their location in time.\label{cap:evolution}}
\end{figure}

Lets first rewrite the integral \ref{eq:theint} as a sum over $K$ blocks of width
$\delta_{i}$ and Taylor expand the memory kernel:
\begin{eqnarray}
I\left(T,\Delta T\right) & = & \sum_{i=1}^{K}\int_{-\delta_{i}/2}^{+\delta_{i}/2}d\epsilon F\left(\tau_{i}+\epsilon,T\right)\alpha\left(\tau_{i}+\epsilon\right)\\
 & \approx & \sum_{i=1}^{K}\int_{-\delta_{i}/2}^{+\delta_{i}/2}d\epsilon F\left(\tau_{i}+\epsilon,T\right)\nonumber \\
 &  & \times\left\{ \alpha\left(\tau_{i}\right)+\alpha^{\p}\left(\tau_{i}\right)\epsilon+\mathcal{O}\left(\epsilon^{2}\right)\right\} \,.
\label{eq:expansion}
\end{eqnarray}
The lowest order approximation to this integral is \begin{equation}
I^{0}\left(T,\Delta T\right)=\sum_{i=1}^{K}\alpha\left(\tau_{i}\right)\int_{-\delta_{i}/2}^{+\delta_{i}/2}d\epsilon F\left(\tau_{i}+\epsilon,T\right)\,.\label{eq:firstord}\end{equation}
This equation represents the fact that we will take some step size,
$h$, to compute the integral of $F$ over the block, but use some
other variable step size to calculate the integral of the product of $\alpha$
and $F$. For the polynomial kernel, the whole procedure is based
on choosing a block size that grows with increasing distance from
the current time, 
\begin{equation}
\delta_{i}=b\tau_{i}\,,
\label{eq:blockgrow}
\end{equation}
or close to it, as shown in Figure \ref{cap:evolution}. We call $b$
the \emph{block parameter}. This function for the block size gives
a constant value when multiplied by the smoothness \ref{eq:smrat}
of the kernel. Assuming the function $F$ is bounded by 1,  
the highest order error on each block is bounded by 
\begin{equation}
\left| \alpha^{\p}\left(\tau_{i}\right)\int_{-\delta_{i}/2}^{+\delta_{i}/2}d\epsilon F\left(\tau_{i}+\epsilon,T\right)\epsilon\right| 
  \le \frac{p}{4}\frac{b^2}{\tau_i^{p-1}} \,.
\end{equation}
Thus, when choosing the blocks in this way, there is a constant ratio, $pb/4$, of the 
bound on the error to the bound on the integral. The bound on the highest order error of the whole integral is
\begin{equation}
\left|\sum_{i=1}^{K}\alpha^{\p}\left(\tau_{i}\right)\int_{-\delta_{i}/2}^{+\delta_{i}/2}d\epsilon F\left(\tau_{i}+\epsilon,T\right)\epsilon\right| 
  \le \frac{pb}{4(p-1)}\frac{1}{\Delta T^{p-1}} \, .
\end{equation}
Roughly, we then expect the error to decrease at least linearly in the block
parameter $b$. We discuss the errors in more detail below.

If we perfectly choose the block positions, we can calculate the number
of blocks required to cover the integration region. The first block
has position (its midpoint) 
\begin{equation}
t_{1}=\Delta T+\frac{b}{2}t_{1}=\frac{\Delta T}{1-\frac{b}{2}}
\end{equation}
and given the position of block $i$, block $i+1$ is at 
\begin{equation}
t_{i+1}=t_{i}+\frac{b}{2}t_{i}+\frac{b}{2}t_{i+1}=\frac{1+\frac{b}{2}}{1-\frac{b}{2}}t_{i}\,.
\end{equation}
The $n^{th}$ block is then at 
\begin{equation}
t_{n}
    = \frac{\left(1+\frac{b}{2}\right)^{n-1}}{\left(1-\frac{b}{2}\right)^{n}}\Delta T\,.
\end{equation}
Since $T$ is arbitrary, the $n^{th}$ block does not necessarily
cover the remaining area. But approximately, 
\begin{eqnarray}
t_{n} & \approx & T-\frac{b}{2}T+\mathcal{O}\left(b^{2}\right)\,.
\end{eqnarray}
and the total number of blocks for the integral is 
\begin{equation}
K=\left\lceil n\right\rceil =\left\lceil \frac{\ln\left(\frac{\Delta T}{T}\right)}{\ln\left(1-\frac{b}{2}\right)-\ln\left(1+\frac{b}{2}\right)}\right\rceil \,.
\label{eq:optK}
\end{equation}
In practice, $K$
will always be bigger than \ref{eq:optK} because of two
factors. One is the finite step size, $h$, which forces the blocks
to be an integral multiple of $h$. Two is that we are not able to
take the complete integral and divide it up perfectly. The division
into blocks has to happen as we solve the integro-differential equation.
We will see in subsection \ref{sub:growing} that neither of these
factors poses a serious problem and the true value of $K$ will be
of the same order as \ref{eq:optK}. For small $b$ (or large
$K$), we can simplify to 
\begin{equation}
K  \approx  \frac{1}{b}\ln\left(\frac{T}{\Delta T}\right)\,.
\end{equation}

Within the algorithm, then, we need to keep track of the $K$ integrals
\begin{equation}
I_{i}=\int_{-\delta_{i}/2}^{+\delta_{i}/2}d\epsilon F\left(\tau_{i}+\epsilon,T\right)\,,\,\delta_{i}=b\tau_{i}
\label{eq:firstordint}
\end{equation}
which can be summed up with the weight $\alpha\left(\tau_{i}\right)$
to get the full integral. Putting in the explicit form of $F$, the
$K$ integrals are \begin{equation}
\int_{-\delta_{i}/2}^{+\delta_{i}/2}d\epsilon e^{-\mathcal{L}(\tau_{i}+\epsilon)}\mathcal{K}^{\p}\rho(T-\tau_{i}-\epsilon)\,.\end{equation}
When we increment $T\to T+h$, we first need to fix the block size,
$\delta_{i}\to B_{i}$. Thus, the blocks will no longer be exactly
$b\tau_{i}$ in width. As $T\to T+h$, $\tau_{i}\to\tau_{i}+h$ and
the integrals are easily updated by \begin{eqnarray}
\int_{-B_{i}/2}^{+B_{i}/2}d\epsilon F\left(\tau_{i}+\epsilon,T\right)\qquad\qquad\qquad\nonumber \\
\to e^{-\mathcal{L}(h)}\left\{ \int_{-B_{i}/2}^{+B_{i}/2}d\epsilon F\left(\tau_{i}+\epsilon,T\right)\right\}  & \,.\end{eqnarray}
After evolving, $B_{i}<b\tau_{i}$, which is acceptable since the
smaller the blocks the smaller the error. The block sizes have to
grow eventually or else we will not get the logarithmic coverage.
Each time we increment $T$ we can check whether nearest neighbor
blocks can be grouped. We can group them so long as \begin{equation}
B^{new}\le b\tau^{new}\label{eq:cond}\end{equation}
where $\tau^{new}$ is the midpoint of the new block. This is the
{}``cleansing'' stage of Figure \ref{cap:evolution}, and is discussed
in more detail in subsection \ref{sub:growing}. When the condition
on the new block size is met, we can simply add the blocks together\begin{eqnarray}
\int_{-B^{new}/2}^{+B^{new}/2}d\epsilon F\left(\tau^{new}+\epsilon,T\right)\qquad\qquad\qquad\qquad\qquad\:\nonumber \\
=\int_{-B_{i}/2}^{+B_{i}/2}d\epsilon F\left(\tau_{i}+\epsilon,T\right)+\int_{-B_{i+1}/2}^{+B_{i+1}/2}d\epsilon F\left(\tau_{i+1}+\epsilon,T\right)\,.\end{eqnarray}

To summarize, we defined a logarithmic covering of the integration
region in terms of growing blocks. As we solve the integro-differential
equation, we can evolve the blocks and perform further groupings to
ensure approximately the same covering. This covering allows for a
logarithmic speedup in computing the history and a logarithmic reduction
in the memory necessary to store the history. At the same time, due
to the increasing smoothness of the polynomial memory kernel, it has
a controllable error. The algorithm can be extended to higher orders as 
shown in \ref{sec:high}.

\subsection{Growing blocks in real time}
\label{sub:growing}

The above description of the blocking algorithm presented the idea
of using a logarithmic covering of the history integral. The formula
\ref{eq:optK} gives the number of blocks needed for the covering
if we could perfectly choose their positions and widths. However,
when solving the integro-differential equation, growing pains exist: the blocks have to
be formed from the smaller blocks of the integration step size $h$.
The block covering will thus be suboptimal.

\begin{figure}
\begin{center}
\includegraphics[width=8cm]{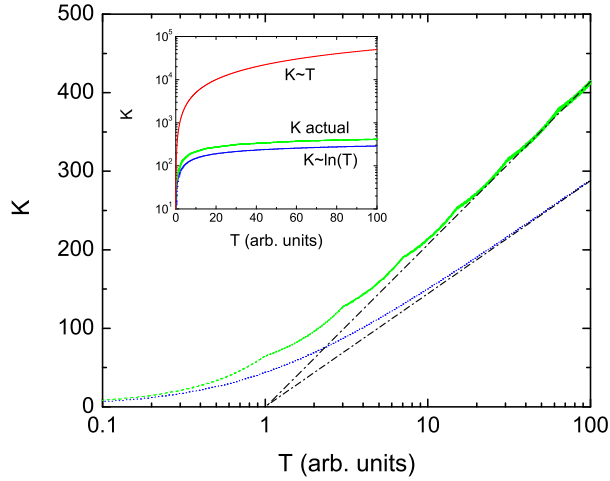}
\end{center}
\caption{The number of blocks $K$ versus $T$. The dotted (blue) curve shows the
optimal case given by \ref{eq:optK}. The dashed (green) curve shows the number of blocks in practice.
The dot-dashed curves represent the asymptotic behaviour of the number
of blocks versus time. These curves intersect the axis at $T=1$
because of the cut-off $\Delta T=1$. 
The inset shows the two latter quantities plotted together with the 
linear growth of $K$ for a typical numerical solution.  
The parameters in the evolution are
the following: $\Delta T=1$, $d=0.016$, and $h=0.002$.\label{fig:kvstime}}
\end{figure}

To determine how close to optimal the number of blocks will be, we
perform a simulation. Here we choose a polynomially decaying function
but with a more natural cut-off, i.e., \begin{equation}
\alpha\left(t\right)=\frac{1}{\left(t+1\right)^{p}}\,.\end{equation}
This will be fully equivalent to $1/\left(t^{\p}\right)^{p}$ with
a cut-off $\Delta T=1$ and a shifted time $t^{\p}=t+1$. Because
of this, the optimal formula \ref{eq:optK} will only hold for large
$T$ (or with $\Delta T=1$ and $T$ replaced with $T+1$). Within
the simulation, we start at $T=0$ and create blocks of size $h$. 
At each step we check if neighboring blocks can be grouped by
satisfying the condition \ref{eq:cond}. If they satisfy that condition,
we group them and check the again with the next block. We perform
this grouping from the smallest to the largest blocks, but the directionality of 
growth does not matter. 
Figure \ref{fig:kvstime} shows how the number of
blocks grow in practice. Although suboptimal, it is still on the same
order as the optimal $K$. How close it is to optimal is dependent
on the step size and block parameter.

\subsection{Errors}
\label{sub:Errors}

Given the blocking construction above, we can analyze how we expect
the errors to behave versus both the number of blocks $K$ and the
computational cost. The first question that is natural to ask is what
error do we make by the replacement of the integral \ref{eq:theint}
with the approximation \ref{eq:firstord}. This gives us an idea of
the error incurred by keeping $K$ blocks of history in a logarithmic
covering.

To answer this first we consider just the integral
\begin{equation}
\int_{0}^{T_{f}}\frac{e^{\imath\omega t}}{\left(t+1\right)^{p}}dt=\int_{\Delta T=1}^{T_{f}+1}\frac{e^{\imath\omega\left(t-1\right)}}{t^{p}}dt
\label{eq:prodint}
\end{equation}
where we have some frequency of oscillation $\omega$ and some power
of the polynomial decay $p$. In the integration below we take $p=2$.
It is important to note that in demonstrating the algorithm with just
the integral \ref{eq:prodint}, and not an integro-differential equation, 
it is not possible to show the computational
speedup. However, it can show the accuracy of the integration as a
function of the logarithmic coverage of the integration region when
using the algorithm. We can also use it to get an idea of how the
error changes when we vary $\omega$ or $p$ without having to solve
the more complicated integro-differential equations. The form
of this integral was chosen because of its relevance to the real-time
simulation of quantum systems, where the integral of the integro-differential
equation will be a sum over oscillating terms times a polynomial memory
kernel.

We gave a simple error bound above. To get a better idea of errors, we can examine the 
behaviour of the integral \ref{eq:prodint} as a function of the blocking parameter, frequency, 
and the integration range (e.g., simulation time). If one takes the highest order error of the 
expansion \ref{eq:expansion}, the worse case error in a block will occur when there is approximately one oscillation in 
the block,
\begin{equation}
b \tau_i = \frac{2 \pi}{\omega} \, ,
\end{equation}
as this is when the asymmetry contributes most. However, this contribution will only occur in some finite region 
of the integral because the block sizes vary. When one changes the block parameter, this error will be shifted further back in time, 
e.g., to the left in Figure \ref{cap:evolution}.  
The memory kernel has a smaller value at this past time, and thus the error will be decreased. If one has many frequencies 
contributing to the integrand, then each of their error contributions will be shifted to a later time as one 
decreases the block parameter. Thus, the error will be systematically reduced.
The same conclusion can be drawn if one expands both the functions in the 
integrand around their midpoint for each block. The fractional error (the highest 
order error term as a fraction of the zeroth order term) in this case is
\begin{equation}
\frac{b^2}{24}\left( p(p+1)- 2 \imath \omega p \tau_i - \omega^2 \tau_i^2 \right) \, .
\label{eq:errexp}
\end{equation}
The expansion will be invalid at approximately
\begin{equation}
\frac{\omega^2 b^2 \tau_i^2}{24}\approx 1 \, ,
\end{equation}
which gives $b \tau_i \sim 5/\omega$.

We compare the logarithmic covering to the midpoint method with step
sizes $h=10^{-3}*2^{j}$ with $j=1,\ldots,9$. This gives a number
of blocks $K_{mid}=T_{f}/h$. For the algorithm, we use a step size
$h=10^{-3}$ to compute the integrals \ref{eq:firstordint} and then
use block parameters $b=10^{-3}*2^{j}$ with $j=1,\ldots,8$ to compute
\ref{eq:firstord}. The number of blocks $K_{alg}$ is then given
approximately by \ref{eq:optK}. It is only approximate because
each block width has to be rounded down to a multiple of $h$, this
gives a larger number of blocks. We use this larger number of blocks
as the value for $K_{alg}$.

If we are just looking at the integral \ref{eq:prodint}, we expect
the error due to the algorithm to be 
\begin{equation}
E_{alg} \propto  K_{alg} b^{3} \approx \frac{1}{K_{alg}^{2}}\left(\ln\frac{T_{f}}{\Delta T}\right)^{3}
\end{equation}
so long as $b T \lessapprox 2 \pi /\omega$.
This can be compared to the error of the midpoint method \begin{equation}
E_{mid}\propto K_{mid} h^3 \approx \frac{T_{f}^{3}}{K_{mid}^{2}}\,.\end{equation}
Thus, if we want a fixed error, $E$, and end time, we get a difference
in the number of blocks of \begin{equation}
K_{mid}\propto\left(\frac{T_{f}^{3}}{E}\right)^{1/2}\end{equation}
compared to \begin{equation}
K_{alg}\propto\left(\frac{\left(\ln\frac{T_{f}}{\Delta T}\right)^{3}}{E}\right)^{1/2}\,,\end{equation}
which is quite large for long times.

\begin{figure}
\begin{center}
\includegraphics[width=8cm]{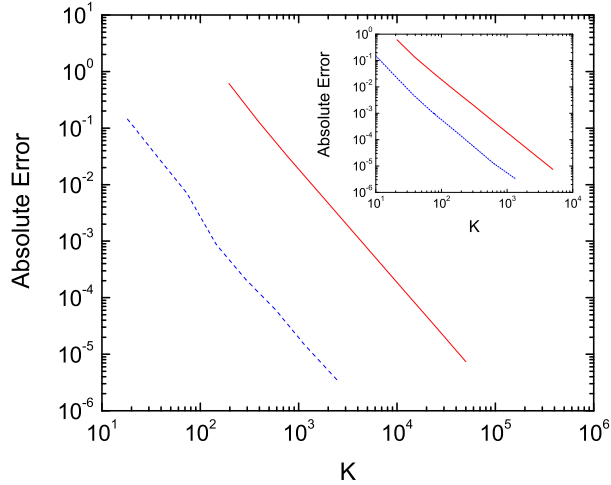}
\end{center}
\caption{Absolute error versus number of blocks for the two methods of calculating
the integral \ref{eq:prodint} with $\omega=2\pi$ and $p=2$ (similar
curves were found for other $\omega$). The solid (red) curves are for the
midpoint method and the dashed (blue) curves are for the first order blocking
algorithm. The main figure shows the error for $T_{f}=100$ and the
inset shows it for $T_{f}=10$. Since the midpoint method is second
order, we find what is expected, $E\propto K^{-2}$. The 
blocking algorithm has the same dependence on $K$.
The difference in error for the same number of blocks with the two
methods is dependent on $T/\ln T$,
reflecting the fact that the algorithm chooses blocks at optimal places
and only covers the integration area with $\ln T$ blocks.
We take as the exact value the one computed with a step size $h=10^{-4}$.
\label{fig:errorvsblocks}}
\end{figure}

Figure \ref{fig:errorvsblocks} shows how the error behaves versus
the number of blocks used to compute the integral. The fast decrease of error as a 
function of $K_{alg}$ shows
that we can block together integrals as we evolve the integro-differential
equation and have a controlled error. Further, the longer the simulation
time, $T_{f}$, the better performing the algorithm should be when compared to 
the midpoint method. In
addition, the overall performance of the algorithm should not be significantly
affected by the frequency of integrand oscillation $\omega$ or the
power $p$. One interesting feature of the error versus $K_{alg}$ figure 
is the slight 
dip in the error at approximately $bT\sim 2 \pi / \omega$, which represents the 
error discussed above being pushed out of the integration region.

The second question we can ask is how the error depends on the computational
cost compared to more traditional methods. If we use the method
for integro-differential equations outlined in \ref{sec:rk}, we have
an error at each step of $h^{3}$ and we have $N$ steps. Thus 
\begin{equation}
E_{mid} \propto N h^{3} \propto \frac{1}{C_{pu}} \, ,
\end{equation}
where $C_{pu}\propto N^{2}$ is the computational cost and we hold the simulation time fixed. 
For the algorithm 
\begin{equation}
E_{alg} \propto \frac{1}{C_{pu}^2}
\end{equation}
where we hold the step size $h$ fixed and $C_{pu}=NK$. 
That is, the error goes down with a higher power in the computational
cost.

Of course, using the algorithm is not the only way to get the error
to scale better with the computational cost. One can also just block
the history with constant blocks larger than $h$. Although the error does scale better, the error can
never be lower than just using the brute force method, e.g., the error
versus cost curve will never cross the brute force curve. For the
algorithm these two curves do cross, as we will see below.

\section{Examples}
\label{sec:Examples}

We consider two example problems to demonstrate the performance of
the algorithm: \emph{(1)} solving an integro differential equation
that has both an oscillating function and a polynomially decaying
function in the integral and \emph{(2)} the NIBA equations for the spin-boson model.

\subsection{Oscillating integro-differential equation}

\begin{figure}
\begin{center}
\includegraphics[width=8cm]{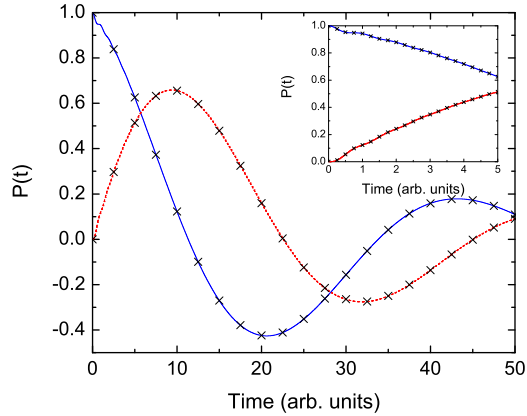}
\end{center}
\caption{Real time evolution of \ref{eq:osceq} with $\omega=2\pi$.
The solid (blue) curve shows the real part of $P\left(t\right)$ and the dashed (red) 
curve the imaginary part. These curves were calculate with the algorithm
with $h=0.002$ and $d=0.016$. The crosses represent 
the solution using the midpoint procedure of \ref{sec:rk} 
with a step size of $h=0.001$. The inset shows a blow up of the
initial time where there is some transient structure in the solution.\label{fig:osc}}
\end{figure}

The oscillating integro-differential equation we use is 
\begin{equation}
\dot{n}\left(t\right)=\imath\omega n\left(t\right)-\int_{0}^{t}dt^{\p}\alpha\left(t-t^{\p}\right)n\left(t^{\p}\right)\end{equation}
with \begin{equation}
\alpha(t)=\frac{1}{\left(t+1\right)^{2}}\,.\label{eq:expoly}\end{equation}
We get rid of the oscillation in the differential equation by setting
\begin{equation}
P\left(t\right)=e^{-\imath\omega t}n\left(t\right)\,,\end{equation}
which gives 
\begin{equation}
\dot{P}\left(t\right)=-\int_{0}^{t}dt^{\p}\alpha\left(t-t^{\p}\right)e^{-\imath\omega\left(t-t^{\p}\right)}P\left(t^{\p}\right)\,.
\label{eq:osceq}
\end{equation}
This integro-differential equation involves the computation of an
integral similar to \ref{eq:prodint}, with $p=2$, at each
time step. With the algorithm, we will be able to recycle the integral
with the logarithmic covering at each step, rather than computing
it from scratch. Note that we choose blocks by $b\left(t+1\right)$
here because we have the cut-off included in the polynomial decay.
The simulation results for one set of parameters are shown in Figure
\ref{fig:osc}.

\begin{figure}
\begin{center}
\includegraphics[width=8cm]{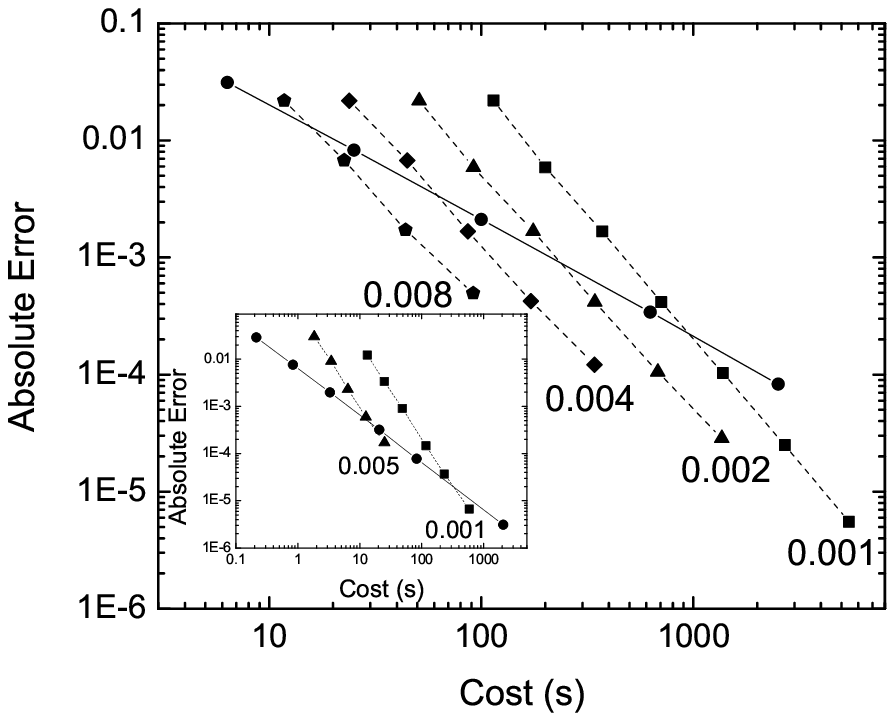}
\end{center}
\caption{Absolute error versus computational cost. The error is computed by
comparing the simulation results with those the brute force method
with $h=0.001$ for $T_{f}=50$ (in the inset, the ``exact'' solution
is taken as the brute method with $h=0.0002$ and $T_{f}=10$). Circles 
represent the brute force method for step sizes 0.1, 0.05, 0.025,
0.01, and 0.005 (in the inset, 0.1, 0.05, 0.025, 0.01, 0.005, and
0.001). The other data represents using the algorithm with the step
size as shown and each data point has a different block parameter
starting with $2h$ and increasing by factors of 2. For the longer simulation time of $T_{f}=50$,
the algorithm has quite a substantial decrease in error for the same
computational cost (for the shorter time in the inset, one can see
that there is barely a gain. This is due to the overhead
cost in implementing the algorithm). Also, the curves have the 
behaviour discussed in subsection \ref{sub:Errors}. 
We perform all computations on a P4, 3 GHz processor.} 
\label{fig:ErrorvsCost}
\end{figure}

Now we examine the error versus computational cost.
Figure \ref{fig:ErrorvsCost} is one of the main results of
this paper. It shows three key results: (1) The algorithm has significantly
better error for the same computational cost compared to a brute
force method in a large range of parameter space. (2) As the simulation
time is increased, the efficiency of the
algorithm drastically increases 
compared to the brute force method. (3) The figure also
suggests how to use the whole ansatz. Rather than have two parameters
$b$ and $h$, one should set $b\propto h$ (for polynomial decay).
This will give a similar convergence to the exact answer as the second
order method, but a smaller factor of proportionality.

As mentioned in subsection \ref{sub:Errors}, one can also consider
another method that just uses a constant grouping to reduce the number
of blocks in the history. However, this method can not achieve better
results as determined by the error versus computational cost. For a given step
size, the highest accuracy will be achieved by using the brute procedure.
A lower computational cost can be achieved by grouping the history
into constant blocks, but the reduction in the computational cost
will yield an even stronger loss in accuracy because $E\propto C^{-2}$, 
when one step size is fixed.
Thus, a similar figure to Figure \ref{fig:ErrorvsCost} would show
curves moving only upward from the brute force line and not crossing
it.

\subsection{NIBA equations\label{sub:NIBA-equations}}

The NIBA for the spin-boson model gives a good physical example for
a system with a polynomially decaying kernel. The NIBA equations are
derived by Leggett \etal \cite{leggett}, see also \ref{sec:NIBA} for
a simple derivation,
from which we can get the equation of motion
\begin{equation}
\dot{P}(t)=-\int_{0}^{t}dt^{\p}f(t-t^{\p})P(t^{\p})\end{equation}
where \begin{equation}
f(t)=\Delta^{2}\cos\left[2\alpha\tan^{-1}t\right]\left[\frac{\left(\pi t/\beta\omega_{c}\right)\mathrm{csch}\left(\pi t/\beta\omega_{c}\right)}{\left(1+t^{2}\right)^{1/2}}\right]^{2\alpha}\ .\label{eq:NIBAker}\end{equation}
Time is in units of the cut-off frequency $\omega_{c}^{-1}$. At zero temperature ($\beta\to\infty$)
the kernel becomes 
\begin{equation}
f(t)=\frac{\Delta^{2}\cos\left[2\alpha\tan^{-1}t\right]}{\left(1+t^{2}\right)^{\alpha}} \, .
\end{equation}
The $1+t^{2}$ has a physical cut-off to the low time behaviour
of the polynomial decay.
There are only two parameters of interest to us. We set $\Delta=0.2$
as it is required to be small (see \ref{sec:NIBA}) and we vary the dissipation
strength $\alpha$ to see the behaviour of $P\left(t\right)$ on different sides of
the dissipative transition. Varying $\alpha$ also shows the algorithm at work for different powers of polynomial decay.


Depending on the value of the dissipation strength $\alpha$, we have
to use a different cut-off $\Delta T$. For $\alpha=0.5$, $1.0$, and $1.5$
we use $\Delta T=1$, $3$, and $2$, respectively. After these cut-offs, $f\left(t\right)$
is increasingly smooth (it is also smoother than 
bare polynomial decay). We want to point out that due to the form of
the polynomial decay, the most efficient way to choose block sizes
is by $b\left(1+t^{2}\right)/t$, which just comes from the inverse
of the increasing smoothness. Despite this, we still use the less
efficient $bt$. 

\begin{figure}
\begin{center}
\includegraphics[width=8cm]{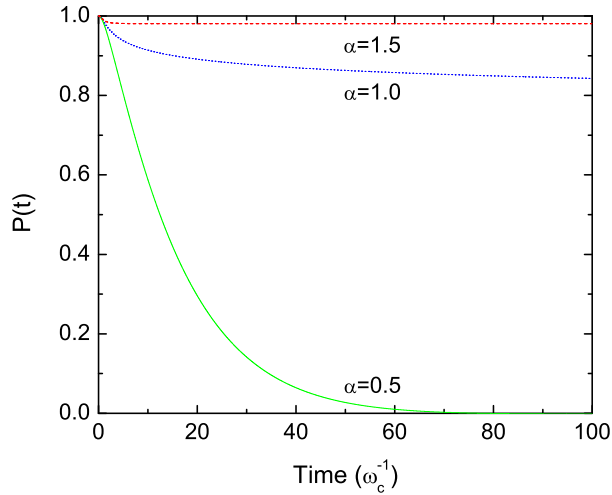}
\end{center}
\caption{NIBA equations simulated in real time for $\alpha=0.5$, $1.0$, and $1.5$.
The simulations were done with $h=0.008$ and $d=0.016$. They 
took approximately two minutes of CPU time on a P4, 3 GHz processor.
This is orders of magnitude faster than a brute force approach would
take to get the same error. From the figure we can clearly observe
the physics of the different regimes. For weak dissipation, one gets
decay to $P=0$. For the transition, $\alpha=1$, one gets $1/t^{2}$
decay of $P\left(t\right)$. Above the transition, $P\left(t\right)$
gets locked into a nonzero value. \label{fig:NIBA}}
\end{figure}

The simulation results for the NIBA equations are shown in Figure
\ref{fig:NIBA}. Using the algorithm allows one to have long simulation
times for negligible computational cost. Further, simulations of NIBA-like 
equations on a lattice or with time dependent Hamiltonians can be 
done efficiently with the method.

We want to emphasize that the simulation of the NIBA equations
at finite temperature \ref{eq:NIBAker} can also use the
algorithm presented in this paper. The finite temperature, however,
introduces a long-time cut-off into the kernel, beyond which the kernel
may not be increasingly smooth. The contribution beyond this cut-off,
though, is negligible. Thus, the algorithm can be used for times less
than $\sim\beta$ exactly how it is used for the zero temperature
case, and there can be a truncation of the kernel at times beyond
times $\sim\beta$.

\section{Conclusions\label{sec:Conclusions}}

We have given an efficient and stable numerical algorithm for solving
integro-differential equations with increasingly smooth memory kernels.
The general computational speedup is $T^{2}\to TC\left(T\right)$,
where $C\left(T\right)$ depends on how fast the kernel gets smooth.
For example, the computational cost of the algorithm for polynomially
decaying kernels is $T\ln T$ rather than the usual $T^{2}$. 

Using a simple integro-differential equation, we demonstrated how well
the algorithm performs compared to a second order, constant step size
method. For long times, there is quite a substantial speedup in the
computation cost to achieve a given error. The solution to the NIBA
equations for the spin-boson model in real-time showed that one can
get results and analyze the situation quite rapidly. Similar procedures
can be applied to other forms of memory kernels which satisfy the
increasingly smooth condition \ref{eq:incsmooth}. In these other
cases, the computational speedup can be better or worse than the case presented here. 

In practice, the usefulness of this algorithm is due to the possibility of 
its incorporation into other techniques, such as matrix product state algorithms, or 
its use with time-dependent Hamiltonian terms. 
Thus, one can simulate lattices or driven systems 
subject to strong dissipation. The algorithm can also be used for 
equations less restrictive than \ref{eq:integro}, such
as integro-differential equations with memory kernels dependent on
both time arguments.

\ack

The author would like to thank G. Vidal and G. Refael for helpful
discussions. This research was supported by the National Science Foundation
through its Graduate Fellowship program and a Sigma Xi Grant-in-Aid
of Research.

\appendix
\setcounter{section}{0}

\section{Higher order blocking algorithms}
\label{sec:high}

The algorithm above can be extended to higher orders. For instance, if we want the
next order method, we need to be able to evaluate and update \begin{eqnarray}
I^{1}\left(T,\Delta T\right)\qquad\qquad\qquad\qquad\qquad\qquad\qquad\qquad\quad\nonumber \\
=I^{0}\left(T,\Delta T\right)+\sum_{i=1}^{K}\alpha^{\p}\left(\tau_{i}\right)\int_{-\delta_{i}/2}^{+\delta_{i}/2}d\epsilon F\left(\tau_{i}+\epsilon,T\right)\epsilon\,.\end{eqnarray}
The latter term picks out the asymmetric part of $F$ inside of block
$i$ and then multiplies it by the derivative of $\alpha$. Lets define
this asymmetric part of the block as \begin{equation}
A_{i}\left(T\right)\equiv\int_{-\delta_{i}/2}^{+\delta_{i}/2}d\epsilon F\left(\tau_{i}+\epsilon,T\right)\epsilon\,.\end{equation}
If we suppose we have these asymmetric integrals at a given time,
and we want to increment $T$, we first need to fix the block size,
as before. Then we can update them by \begin{equation}
A_{i}\left(T+h\right)=e^{-\mathcal{L}(h)}A_{i}\left(T\right)\end{equation}
We also need to be able to add these blocks together. We can do this
by \begin{eqnarray}
A^{new}\left(T\right) & = & \int_{-B^{new}/2}^{+B^{new}/2}d\epsilon F\left(\tau^{new}+\epsilon,T\right)\epsilon\\
 & = & A_{i}\left(T\right)-B_{i+1}I_{i}\left(T\right)+A_{i+1}\left(T\right)+B_{i}I_{i+1}\left(T\right)\nonumber \end{eqnarray}
where we use the asymmetric integrals from before we did the blocking
and also the first order integrals \ref{eq:firstordint} for the two
blocks. The error for an order $z$ algorithm will have a bound proportional to $b^z$


\section{Two stage Runge-Kutta method}
\label{sec:rk}

In this article, we compare the proposed numerical ansatz to a two stage Runge-Kutta 
method. Since we are dealing with integro-differential
equations, we give the details of the second order technique we use.
In the main text, we discuss how the errors scale with total simulation
time, step size, and computational cost.

For all the examples in this work, the basic integro-differential equation \ref{eq:integro}, 
is reduced to 
\begin{eqnarray}
\pd{\rho(t)}{t} & = & \int_{0}^{t}dt^{\p}\alpha\left(t-t^{\p}\right)e^{-\mathcal{L}\left(t-t^{\p}\right)}\rho\left(t^{\p}\right)\,,
\end{eqnarray}
e.g., $\mathcal{K},\mathcal{K}^{\p}=\mathrm{I}$. 
In a generic form we can write 
\begin{equation}
\pd{\rho(t)}{t}=f\left[t,\rho\left(t\right),\int_{0}^{t}dt^{\p}\mathcal{F}\left(t-t^{\p}\right)\rho\left(t^{\p}\right)\right]\,.
\end{equation}
Discretizing time as \begin{equation}
t_{n}=t_{0}+nh\,,\end{equation}
we can write a general two stage Runge-Kutta integration scheme

\begin{equation}
\rho_{n+1}=\rho_{n}+h\, d\rho_{n}\,,\end{equation}

\begin{equation}
d\rho_{n}=f\left[t,P_{n},\tilde{z}_{n}\right]\,,\end{equation}

\begin{equation}
P_{n}=\rho_{n}+dP_{n}\,,\end{equation}
\begin{equation}
dP_{n}=\frac{h}{2}f\left[t,\rho_{n},z_{n}\right]\,,\end{equation}
\begin{equation}
z_{n}=h\sum_{m=1}^{n-1}\bar{\mathcal{F}}_{nm}\left(\rho_{m}+\rho_{m+1}\right)/2\,,\end{equation}
and\begin{eqnarray}
\tilde{z}_{n} & = & h\sum_{m=1}^{n-1}\bar{\mathcal{F}}_{nm}\left(\rho_{m}+\rho_{m+1}\right)/2\nonumber \\
 &  & +\frac{h}{2}\bar{\mathcal{F}}_{0}\left(\rho_{n}+P_{n}\right)/2\,,\end{eqnarray}
where \begin{equation}
\bar{\mathcal{F}}_{nm}=\frac{1}{h}\avg{\mathcal{F}\left(t_{n}-t^{\p}\right)}_{t_{m}}^{t_{m}+h}=\frac{1}{h}\int_{t_{m}}^{t_{m}+h}dt^{\p}\mathcal{F}\left(t_{n}-t^{\p}\right)\end{equation}
and \begin{equation}
\bar{\mathcal{F}}_{0}=\frac{2}{h}\int_{t_{n}-h/2}^{t_{n}}dt^{\p}\mathcal{F}\left(t_{n}-t^{\p}\right)\,.\end{equation}
Although using the average $\bar{\mathcal{F}}_{nm}$ over an interval
does not increase the order of the method, it does preserve important
properties of the kernel such as its total integration. This is very important 
in cases where the kernel integrates to zero and thus the transient behaviour completely determines 
the steady state. We use the average of the kernel over each block with the algorithm as well.
The Runge-Kutta scheme can of course be generalized to higher stages and to kernels
with two time arguments.\cite{Leathers,brunner,Linz,lubich}

\section{Simple derivation of the NIBA equations}
\label{sec:NIBA}

In this appendix we give a simple derivation of the NIBA equations
for the spin-boson model to show how polynomially decaying memory
kernels can arise physically in the case of strong dissipation. The derivation
of the NIBA equations is based on the observation that they come from
a Born approximation of a transformed Hamiltonian.\cite{Aslangul}

The spin-boson Hamiltonian is \begin{equation}
H_{SB}=-\frac{1}{2}\Delta\sigma_{x}+\frac{1}{2}\epsilon\sigma_{z}+\sum_{k}\omega_{k}a_{k}^{\dagger}a_{k}+\gamma\sigma_{z}\sum_{k}g_{k}\left(a_{k}^{\dagger}+a_{k}\right)\end{equation}
where we have a two level system with internal coupling constant $\Delta$
and bias $\epsilon$. The two level system is coupled to an collection
of bosons of frequencies $\left\{ \omega_{k}\right\} $ with a coupling
constant $g_{k}=c_{k}/\sqrt{2m_{k}\omega_{k}}$ and overall coupling factor
$\gamma$. The spectral density of the bath is given by \begin{equation}
J\left(\omega\right)=\pi\sum_{k}g_{k}^{2}\delta\left(\omega-\omega_{k}\right)\,.\end{equation}

The physical scenario we want to explore is one in which the two level
system is initially held fixed in an eigenstate of $\sigma_{z}$ while the bath
equilibrates around this state. That is, the bath will equilibrate
around the Hamiltonian \begin{equation}
H_{R}=\sum_{k}\omega_{k}a_{k}^{\dagger}a_{k}+\sum_{k}\gamma g_{k}\left(a_{k}^{\dagger}+a_{k}\right)\end{equation}
if we hold the two level system in the $+1$ eigenstate of $\sigma_{z}$.
This gives the thermal starting state for the bath \begin{equation}
R_{0}=\frac{e^{-\beta H_{R}}}{Z_{R}}\,.\end{equation}
 Then, we want to release the two level system and follow its dynamics
in real time. In particular, we look at the expectation value of $\sigma_{z}$\begin{eqnarray}
P\left(t\right) & \equiv & \avg{\sigma_{z}}_{t}\\
 & = & \tr\left\{ e^{\imath H_{SB}t}\sigma_{z}e^{-\imath H_{SB}t}\proj{1}{1}\otimes R_{0}\right\} \, .
\end{eqnarray}

Since we are interested in strong dissipation, we can perform a canonical
transformation on this Hamiltonian to incorporate all orders of the
system-bath interaction. With \begin{equation}
S=-\sum_{k}\frac{\gamma g_{k}}{\omega_{k}}\left(a_{k}-a_{k}^{\dagger}\right)\sigma_{z}\end{equation}
we get \begin{eqnarray}
H & = & e^{S}H_{SB}e^{-S}\\
 & = & -\frac{1}{2}\Delta\left(\sigma_{+}B_{-}+\sigma_{-}B_{+}\right)+\frac{1}{2}\epsilon\sigma_{z}+\sum_{k}\omega_{k}a_{k}^{\dagger}a_{k}\end{eqnarray}
where \begin{equation}
B_{\pm}=\exp\left\{ \pm2\sum_{k}\frac{\gamma g_{k}}{\omega_{k}}\left(a_{k}-a_{k}^{\dg}\right)\right\} \,.\end{equation}
For the unbiased case, $\epsilon=0$, this gives the interaction picture
Hamiltonian \begin{equation}
H^{I}\left(t\right)=-\frac{1}{2}\Delta\left(\sigma_{+}B_{-}^{I}\left(t\right)+\sigma_{-}B_{+}^{I}\left(t\right)\right)\end{equation}
with \begin{equation}
B_{\pm}^{I}\left(t\right)=\exp\left\{ \pm2\sum_{k}\frac{\gamma g_{k}}{\omega_{k}}\left(a_{k}e^{-\imath\omega_{k}t}-a_{k}^{\dg}e^{\imath\omega_{k}t}\right)\right\} \,.\end{equation}
We can then transform the equation for $P\left(t\right)$ to get \begin{eqnarray}
P\left(t\right) & = & \tr\left\{ e^{\imath Ht}\sigma_{z}e^{-\imath Ht}\proj{1}{1}\otimes R_{0}^{\p}\right\} \label{eq:poft}\end{eqnarray}
where \begin{equation}
R_{0}^{\p}=e^{-\beta\sum_{k}\omega_{k}a_{k}^{\dg}a_{k}}/Z_{R}^{\p}\,.\end{equation}

We can find the master equation in the Born approximation for $P\left(t\right)$,
also known as the noninteracting blip approximation, by performing
perturbation theory on \ref{eq:poft}. To second order in
$\Delta$, \begin{equation}
\dot{P}(t)=-\int_{0}^{t}dt^{\p}f(t-t^{\p})P(t^{\p})\end{equation}
 with \begin{eqnarray}
f\left(t\right) & = & \frac{\Delta^{2}}{4}\left\{ \avg{\left[B_{+}^{I}\left(t\right),B_{-}^{I}\left(0\right)\right]}_{R_{0}^{\p}}\right.\nonumber \\
 &  & \left.+\avg{\left[B_{+}^{I}\left(-t\right),B_{-}^{I}\left(0\right)\right]}_{R_{0}^{\p}}\right\} \\
 & = & \frac{\Delta^{2}}{2}\avg{\left[B_{+}^{I}\left(t\right),B_{-}^{I}\left(0\right)\right]}_{R_{0}^{\p}}\end{eqnarray}
where we have used that the correlation functions are equal. To compute
$f\left(t\right)$ we can use the Feynman disentangling of operators
\cite{Mahan} to get, in the notation of Leggett \etal \cite{leggett},
\begin{equation}
f\left(t\right)=\Delta^{2}\cos\left\{ \frac{4\gamma^{2}}{\pi}Q_{1}\left(t\right)\right\} \exp\left\{ -\frac{4\gamma^{2}}{\pi}Q_{2}\left(t\right)\right\}
\end{equation}
 with \begin{equation}
Q_{1}\left(t\right)=\int_{0}^{\infty}d\omega\frac{J\left(\omega\right)}{\omega^{2}}\sin\left(\omega t\right)\end{equation}
and \begin{equation}
Q_{2}\left(t\right)=\int_{0}^{\infty}d\omega\frac{J\left(\omega\right)}{\omega^{2}}\left(1-\cos\omega t\right)\coth\left(\frac{\beta\omega}{2}\right)\,.\end{equation}
For ohmic dissipation, $J\left(\omega\right)=\eta\omega\exp\left(-\omega/\omega_{c}\right)$,
these quantities become \begin{equation}
Q_{1}\left(t\right)=\eta\tan^{-1}\omega_{c}t\end{equation}
and \begin{equation}
Q_{2}\left(t\right)=\frac{\eta}{2}\ln\left(1+\omega_{c}^{2}t^{2}\right)+\eta\ln\left(\frac{\beta}{\pi t}\sinh\frac{\pi t}{\beta}\right)\,.\end{equation}
With $\alpha\equiv2\eta\gamma^{2}/\pi$, this gives \ref{eq:NIBAker}
for the NIBA memory kernel.

\end{document}